\documentclass[%
 reprint,
showpacs,
amsmath,amssymb,
 aps,
]{revtex4-1}

\usepackage{fancyhdr}
\usepackage{graphicx}
\usepackage{dcolumn}
\usepackage{bm}

\begin{document}
\title{Bright Emission from Dark Sources in Hyperbolic Media}
\author{Evgenii E. Narimanov}
\affiliation{School of Electrical and Computer Engineering and Birck Nanotechnology Center, Purdue University, West Lafayette, Indiana 47907, USA}
\date{\today}

\begin{abstract}
Hyperbolic media enable ultra-strong light--matter interactions through their  extreme field localization and small mode volumes, but low-loss realizations are fundamentally limited to the mid-infrared, owing to the long lifetimes of optical phonons in high-quality crystals. Here we show that bright emitters operating at visible or near-infrared frequencies can be used to generate radiation in this regime by inducing mid-infrared population dynamics, thereby creating a source in the hyperbolic frequency band without a corresponding dipole transition. We demonstrate that even a source with vanishing dipole and higher multipole moments -- strictly non-radiating in any isotropic medium -- becomes radiatively active in a hyperbolic environment. This enables visible and near-infrared control of light--matter interactions in polaritonic hyperbolic materials, establishing a new low-loss solid-state quantum optics platform.
\end{abstract}

\maketitle

Hyperbolic media have emerged as a powerful platform for controlling light--matter interaction at the nanoscale.\cite{HQP} Their highly anisotropic dispersion enables an exceptionally large photonic density of states,\cite{PDOSprl} extreme field confinement,\cite{PDOSapl,Vinod} and propagation of high-momentum modes that are inaccessible in conventional materials.\cite{PN,hyperlens1,hyperlens2} These properties have led to a wide range of proposed applications, including enhanced spontaneous emission, strong coupling at deeply subwavelength scales, and long-range, highly directional energy transfer.\cite{Vinod,hyperlensE1,hyperlensE2} In this context, hyperbolic systems appear ideally suited for quantum optics and quantum information processing, where strong and controllable interactions between emitters are essential.

However, this promise is fundamentally limited by material loss. Most realizations of hyperbolic response at optical frequencies rely on electronic excitations, which inevitably introduce significant dissipation.\cite{Noginov,Zhang} The same mechanisms that enable strong confinement also lead to rapid decay of electromagnetic energy, severely restricting coherence times and interaction lengths.\cite{ScienceTT} As a result, despite their appealing theoretical properties, conventional hyperbolic platforms are often unsuitable for quantum applications where low loss is critical.

A promising route to mitigate these losses has emerged in polar dielectric materials operating within their Reststrahlen bands.\cite{Bi,hBN1,DaiBasov2015} In these systems, hyperbolic dispersion is supported by optical phonons rather than electronic excitations, resulting in substantially longer lifetimes and reduced dissipation.\cite{BasonFoglerDeAbajo2016,WangLow2024} Materials such as hexagonal boron nitride ($h$BN) have demonstrated low-loss propagation of hyperbolic 
phonon--polaritons,\cite{hBN2,hBN3,hBN4,YoxallHillenbrand2015,LowKoppens2017,NiBasov2021,SternbachBasov2021,SternbachBasov2023,JiaAsgari2022} with coherence properties that can be further improved through isotopic purification.\cite{hBNisotope} These advances suggest that phonon-based hyperbolic media may provide a viable path toward low-loss nanophotonic platforms.

Yet, this approach introduces a different and equally fundamental limitation. The hyperbolic response in such materials is restricted to mid-infrared frequencies defined by the phonon spectrum. In contrast, the most robust and well-controlled quantum emitters---such as atomic systems, molecules, and solid-state defects---operate primarily in the visible and near-infrared.\cite{Kimble2008} This spectral mismatch severely constrains the direct use of low-loss hyperbolic media for quantum applications, as the available emitters and the optimal material platforms reside in fundamentally different frequency ranges.

This mismatch leads to a fundamental impasse. One possibility is to identify or engineer quantum emitters that operate directly in the mid-infrared; however, despite decades of effort, no class of emitters with the coherence, brightness, and controllability required for quantum applications has emerged in this spectral range. Alternatively, one might attempt to extend hyperbolic response into the visible or near-infrared by relying on electronic excitations. Yet this inevitably reintroduces strong material losses, undermining the very advantage that makes phonon-based systems attractive. 

As a result, the field is confronted with a seemingly unavoidable tradeoff: either operate in a frequency range where suitable emitters are unavailable, or accept levels of dissipation that preclude coherent quantum control.

Here we show how to resolve this impasse. Rather than searching for new mid-infrared emitters, we show that population dynamics of conventional visible and near-infrared emitters can become radiatively active in the hyperbolic band despite the absence of a corresponding transition in the emitter. The foundation of our approach is the well-known fact that a high-quality
quantum emitter can be driven coherently so that its internal dynamics acquire
frequency components far from its natural resonance, for example through strong
off-resonant drive or multifrequency excitation.\cite{Boyd} In particular, optical driving fields can generate responses at mid-infrared frequencies through beating and related mechanisms. However, for a spherically symmetric, charge-neutral emitter, this does not by itself lead to radiation: the resulting charge distribution remains symmetric and carries neither monopole nor dipole nor any multipole moment, and therefore would normally be expected to remain dark.\cite{Boyd} In an isotropic medium, this expectation is indeed correct, and the presence of mid-infrared population dynamics does not produce emission.

The central result of this work is that this conclusion fails in an anisotropic medium -- where a driven, spherically symmetric emitter becomes an efficient source of radiation. This effect is fundamentally different from the enhancement of emission from higher-order multipole moments in nanostructures and plasmonic systems,\cite{Rivera2016} as here {\it all} multipole moments of the source vanish identically.

We consider an emitter that is spherically symmetric at all times. Any intrinsic anisotropy would introduce conventional dipolar or higher-order channels and thereby enhance emission. By restricting to the fully isotropic case, the emitter is, in the conventional sense, maximally dark, and any radiation it produces must arise solely from the properties of the surrounding medium.

Under coherent excitation, the population of the emitter's excited state can be strongly modulated in time. This can be achieved using multiple optical fields with controlled frequencies and phases, resulting in a driven dynamics that contains well-defined frequency components in the mid-infrared. This modulation does not alter the spatial symmetry of the emitter:
the charge distribution remains spherically symmetric at all times.
To leading order, the emitter can therefore be described as a time-dependent,
radially symmetric charge distribution---a pulsating sphere---with no dipole,
quadrupole, or higher multipole moment, which in an isotropic medium
produces no radiation.

At distances smaller than the free-space wavelength, the electric field produced by the emitter, can be described within the framework of the 
quasistatic approximation, with the scalar potential 
\begin{eqnarray}
\phi\left({\bf r}, t\right) & = & \int d{\bf k}\int d\omega\ \phi_\omega\left({\bf k}\right) e^{i {\bf k}\cdot{\bf r} - i \omega t}
\end{eqnarray}
defined by the Fourier transform $\rho_\omega\left({\bf k}\right)$  of the effective charge density of the emitter $\rho\left({\bf r}, t\right)$:
\begin{eqnarray}
\phi_\omega\left({\bf k}\right) & = & \frac{4 \pi \rho_\omega\left({\bf k}\right)}{{\bf k}\cdot \boldsymbol{\epsilon}_\omega \cdot {\bf k}}. 
\end{eqnarray}
For an atomic emitter, $\rho\left({\bf r}, t\right)$  is the difference between the charge densities at the excited  and the ground energy levels, multiplied
by the (time-dependent) occupation of the excited state.

The total charge neutrality of the emitter sets its  monopole moment to zero, while spherical symmetry ensures that the dipole and all higher tensor multipoles vanish exactly -- which guarantees that in an isotropic medium the spherical emitter produces no external field. In an anisotropic medium, however, this conclusion no longer holds.

Introducing the second moment of the charge distribution,
\begin{equation}
M_2\left(\omega\right)=\int d^3r\, r^2 \rho_\omega(\mathbf{r}),
\end{equation}
for the scalar potential of the emitter  we obtain
\begin{equation}
\phi_\omega(\mathbf{k})=-\frac{2 \pi}{3}\,\frac{M_2 \ k^2}{\mathbf{k}\cdot \boldsymbol{\epsilon}_\omega\cdot \mathbf{k}}+O(k^4).
\end{equation}

In the isotropic limit, this expression reduces to a spatially localized potential and no external field is generated. In contrast, in an anisotropic medium the field of the spherical emitter extends into the surrounding space, leading to a finite radiated field extending to macroscopic distances.

In Fig.~1 we show the electric field intensity (false color) produced by such an emitter in a hyperbolic medium. The calculation is performed for hexagonal boron nitride (hBN) with its natural material losses. The field exhibits strong directional structure and follows the characteristic hyperbolic cones, in direct analogy with the emission pattern of a conventional dipole in the same medium. This behavior is further enhanced in confined geometries, where boundary conditions lead to electromagnetic field localization, such as the phenomenon of autofocusing in cylindrical geometry.\cite{EN2025}.

\begin{figure*}[htbp] 
   \centering
   \includegraphics[width=6in]{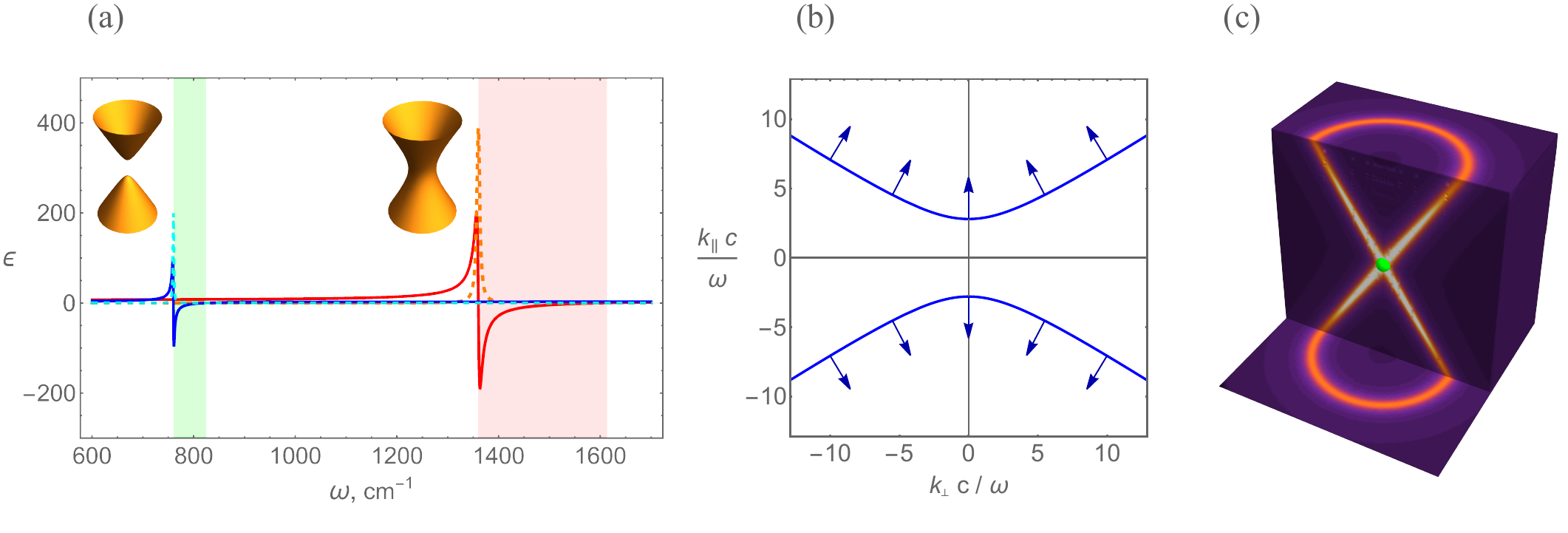} 
      \caption{ Emission from a multipole-free spherical source in a hyperbolic medium.
(a) Frequency-dependent permittivity components of hexagonal boron nitride ($h$BN). The shaded regions indicate spectral bands with hyperbolic dispersion; the insets show the topology of the corresponding iso-frequency surfaces.
(b) Iso-frequency contours in momentum space at free-space wavelength $\lambda = 13\ \mu\mathrm{m}$ ($\omega \approx 769\ \mathrm{cm}^{-1}$), within the lower hyperbolic Reststrahlen band of $h$BN.\cite{hBN1,DaiBasov2015}  The open dispersion curves support propagation of high-momentum modes, with energy flow directed along the hyperbolic cone directions.
(c) Real-space electric field intensity generated by a spherically symmetric, charge-neutral emitter (green dot) with vanishing dipole and higher multipole moments, in an infinite hyperbolic medium. Despite being strictly non-radiating in isotropic media, the source emits strongly along characteristic hyperbolic directions. This behavior arises from the  material anisotropy that opens a radiative channel which does not exist in free space.}
   \label{fig:1emitter}
\end{figure*}

For the interaction energy of two emitters separated by the distance $r$, we obtain
\begin{eqnarray}
U & =& 4 \pi \  {\rm Re} \ \int \frac{d^3k}{(2\pi)^3}\,\frac{ \rho_1(\mathbf{k})\rho^*_2(\mathbf{k})}{\mathbf{k}\cdot \boldsymbol{\epsilon}\cdot \mathbf{k}}
e^{i {\bf k}\cdot{\bf r}},
\end{eqnarray}
which for $r$ much larger than the emitter size $a$ reduces to
\begin{equation}
U(r)= {\rm Re} \ \frac{M_{2,1}M^*_{2,2}}{6 \epsilon_{\mathrm{eff}}(\hat{\mathbf{r}})\,r^5},
\end{equation}
with
\begin{equation}
\epsilon_{\mathrm{eff}}(\hat{\mathbf{r}})=\hat{\mathbf{r}}\cdot \boldsymbol{\epsilon}_{\omega} \cdot \hat{\mathbf{r}},
\end{equation}
where $\hat{\mathbf{r}}=\mathbf{r}/|\mathbf{r}|$.

In a hyperbolic medium, $\epsilon_{\mathrm{eff}}(\hat{\mathbf{r}})$ vanishes along specific directions corresponding to the hyperbolic cones. Along these directions the interaction is strongly enhanced, leading to highly anisotropic and directional coupling between emitters. This behavior is well known for conventional dipole emitters in hyperbolic media, where the interaction follows the same angular dependence but scales as $1/r^3$.

Thus, emitters that are completely dark in an isotropic environment become not only radiatively active but also strongly interacting, with the interaction strength and directionality controlled by the hyperbolic response of the medium.

The anisotropy-enabled interaction channel identified in this work becomes
progressively stronger in confined geometries, where the ``dark-emitter''
interaction described here is enhanced by the phenomenon of autofocusing in
hyperbolic media \cite{EN2025}. In particular, in a cylindrical geometry the
hyperbolic response leads to repeated focusing of the field along the
propagation cones, resulting in strong spatial concentration of energy on the
cylinder axis and  enhanced emitter-emitter interactions.

This effect originates from the directional propagation imposed by hyperbolic dispersion. Energy flows along well-defined characteristic directions, and reflections from the cylindrical boundary repeatedly redirect these trajectories toward the axis, producing a sequence of focal points at
\begin{equation}
z_m = \pm \frac{2 m R}{\mathrm{Re}\,\xi}_\omega, \qquad m=1,2,\ldots,
\end{equation}
where $R$ is the cylinder radius and 
\begin{eqnarray}
\xi_\omega\equiv\sqrt{-\epsilon_{\perp}\left(\omega\right)/\epsilon_{\parallel}\left(\omega\right)}.
\end{eqnarray}
 For typical parameters of hyperbolic phonon-polariton systems, these distances lie in the range of tens to hundreds of nanometers, making the effect directly relevant for nanoscale experiments.

In the presence of material losses, the foci are broadened, with a characteristic axial extent
\begin{equation}
\delta z \sim \ell_f \equiv {R} \ \frac{\mathrm{Im}\,{\xi_\omega}}{\mathrm{Re}\,{\xi_\omega} }.
\end{equation}
At these focal points the field is concentrated into a region of size $\ell_f \ll R$, leading to a strong enhancement of the  emitter-emitter interactions on the axis of a hyperbolic cylinder waveguide. For the corresponding interaction energy as a function of the emitter separation distance $z$, we obtain
\begin{equation}
U(z) = {\rm Re} \ 
\frac{M_{2,1} M^*_{2,2}}{18 \pi   \, a^2 R^3 \epsilon_z}
\sum_{n}
\left[
\frac{\exp\!\left(-\frac{a^2 x_n^2}{4 R^2}\right)}{J_1(x_n)^2}
\right]
K_n\left(z\right)
\end{equation}
where
\begin{equation}
K_n(z) \equiv
\int_{-\infty}^\infty  \, du \ 
\frac{(x_n^2 + u^2)^2}{
u^2 + \frac{\epsilon_z}{\epsilon_\perp}\,x_n^2} \ 
\exp\!\left[i\,\frac{z \, u }{R} -\left(  \frac{a \, u }{2R}\right)^2 \right],
\end{equation}
and $M_{2,n}$ represents the second moment of the charge density of the $n$-th emitter at the frequency $\omega$.

When the emitter separation $z$ is close to the focal distance $z_m$, 
\begin{eqnarray}
z & = & z_m + \delta z, \ \ \left| \delta z\right| \ll R,
\end{eqnarray}
we find
\begin{eqnarray}
U(\delta z) & = & {\rm Re} \ 
\frac{M_{2,1} M^*_{2,2}}{a^2  } \,
\frac{(\epsilon_\perp-\epsilon_z)^2\,\epsilon_\perp}{6\pi \epsilon_z^3}\,
\nonumber \\
& \times & 
\frac{R^2}{
\left[
\delta z \, e^{- \pi m \gamma }
+
i\,2m\gamma\,R \, \sqrt{-\frac{\epsilon_\perp}{\epsilon_z}}\,
\right]^5
},
\end{eqnarray}
where the hyperbolic medium loss tangent
\begin{eqnarray}
\gamma \equiv\frac{ {\rm Im} \xi_\omega }{{\rm Re} \ \xi_\omega }
\end{eqnarray}
so when the emitter separation is close to the focal distance, the interaction energy
\begin{equation}
U \sim U_{\mathrm{f}}\equiv \frac{\left| M_{2}/a \right|^2}{\ell_f^3}\left(\frac{R}{\ell_f}\right)^2
= \frac{\left| M_{2}/a \right|^2}{\ell_f^3} \frac{1}{\gamma^2}
\label{eq:Uf}
\end{equation}

\begin{figure*}[htbp] 
   \centering
   \includegraphics[width=6in]{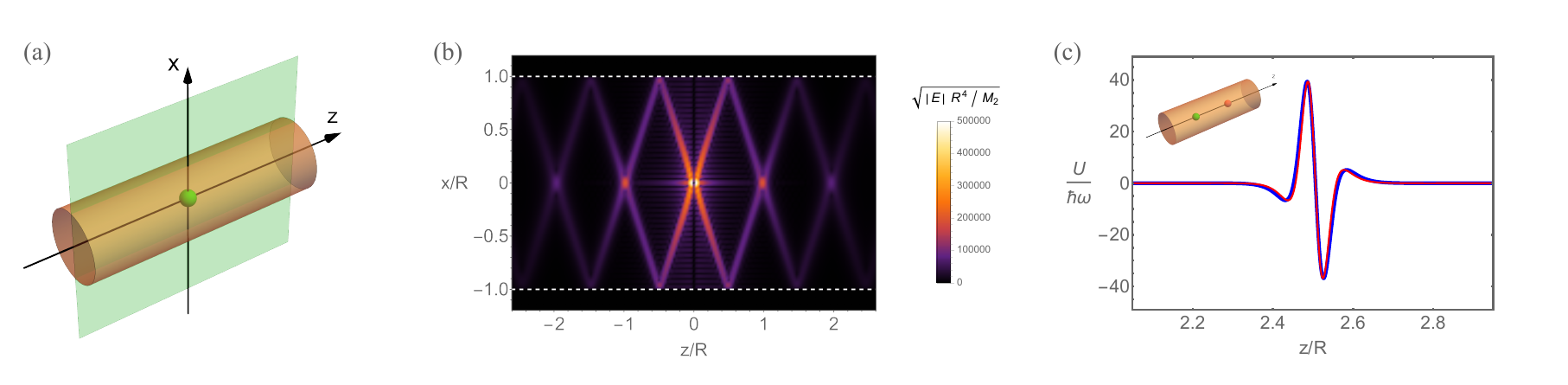}
   \caption{ Autofocusing and interaction of spherically symmetric emitters in a cylindrical hyperbolic medium.
(a) Geometry of the system: a spherically symmetric, charge-neutral emitter (green) is placed on the axis of a cylindrical hyperbolic medium.
(b) Electric field intensity in the $(x,z)$ plane. Hyperbolic propagation in the confined geometry leads to periodic refocusing of the field along the cylinder axis, producing a sequence of localized intensity maxima. The dashed lines indicate the cylinder boundaries.
(c) Interaction energy between two emitters placed on the cylinder axis, as a function of their separation. The refocused field results in strongly enhanced, spatially localized interaction peaks, reflecting the sequence of axial foci. This autofocusing mechanism provides a direct route to strong, directional coupling between otherwise dark emitters.}
   \label{fig:cylinder}
\end{figure*}

For an atomic emitter, $M_2$ is the second moment of the difference between the charge densities
of the excited and the ground states, multiplied by the time-dependent occupation probability of the 
excited state $f_e(t)$, so that with the effective emitter size $a$ we obtain
\begin{eqnarray}
M_2\left(\omega\right) & = & e  a  f_e\left(\omega\right),
\end{eqnarray}
and therefore for two identical emitters the interaction energy (\ref{eq:Uf}) reduces to
\begin{eqnarray}
U(\delta z) & = &   {\rm Re} 
\,
\frac{(\epsilon_\perp-\epsilon_z)^2\,\epsilon_\perp}{6\pi \epsilon_z^3}\, \nonumber \\
& \times & 
\frac{e^2 a^2  R^2 f_e^2}{
\left[
\delta z \, e^{- \pi m \gamma }
+
i\,2m\gamma\,R \, \sqrt{-\frac{\epsilon_\perp}{\epsilon_z}}\,
\right]^5
},
\end{eqnarray}
with the peak values on the order of 
\begin{eqnarray}
U_f \sim \frac{e^2 a^2}{\ell_f^3}  \left(\frac{f_e}{\gamma}\right)^2 
\end{eqnarray}
This behavior is illustrated in Fig. 2(c). 

In the centers of both of the hyperbolic bands in the hexagonal boron nitride ($h$BN),\cite{hBN1,DaiBasov2015} we find $\gamma \sim 0.03$ and 
$\ell \sim R/30$, so that
with atomic-size emitters ($a \sim 1\mathring{\rm A}$) and the cylinder radius $R\sim 100 \ {\rm nm}$, for a modest excited
pump-induced state population $f_e \sim 0.1$, we find 
$U_f \sim 10 \ {\rm meV}$, or $U_f / \hbar\omega  \gtrsim 0.1$ -- so that spherical atomic emitters on the axis of 
a cylinder $h$BN waveguide are strongly coupled at the frequency when the emitter separation matches the focal distance
$2 R / \xi_\omega$. 

Together, Figs. 1 and 2 illustrate complementary manifestations of the same mechanism: anisotropy-enabled propagation of high-
$k$ components in a hyperbolic medium, leading respectively to radiation and to strongly enhanced interactions.

In conclusion, we have shown that a spherically symmetric, charge-neutral emitter,
which is completely dark in an isotropic environment, becomes radiatively active
in an anisotropic medium. This effect does not rely on dipole or higher-multipole
radiation, but arises from an anisotropy-enabled channel associated with the
second moment of the charge distribution. In combination with low-loss
phonon-polariton platforms such as $h$BN, this mechanism provides a direct route
to coupling high-quality quantum emitters in the visible or near-infrared to
mid-infrared hyperbolic modes. Together with autofocusing-enhanced
emitter-emitter interactions, this system offers a viable platform for strong,
directional, and controllable interactions in a new low-loss solid-state
quantum optics platform.

\end{document}